\documentclass{ws06-procs}
\usepackage[latin1]{inputenc}
\begin{document}

\title{Simulation tool of a Supernova search with VST}
%\shorttitle{Instructions for Latex Manuscript}

\author{Rosa Calvi}
%\shortauthors{Lee et al.}
\address{Dip. Scienze Fisiche, Università di Napoli Federico II, via Cinzia 20, 80126 Napoli, Italy
\\INAF- Oss. Astronomico di Capodimonte, Salita Moiariello 16, 80131 Napoli, Italy\\E-mail: calvi@na.astro.it}
\author{Enrico Cappellaro, Maria Teresa Botticella, Marco Riello}
\address{INAF- Oss. Astronomico di Padova, Vic. dell'Osservatorio 5, 35122 Padova, Italy\\
INAF- Oss. Astronomico di Collurania, via M. Maggini, 64100 Teramo, Italy\\European Southern Observatory, K.
Schwarzschild Str.2, 85748 Garching, Germany}

%%%%%%%%%%%%%%%%%%%%%%%%%%%%%%%%%%%%%%%%%%%%%%%%%%%%%%%%%%%%%%
% You may repeat \author \address as often as necessary      %
%%%%%%%%%%%%%%%%%%%%%%%%%%%%%%%%%%%%%%%%%%%%%%%%%%%%%%%%%%%%%%

\maketitle

\abstracts{To improve the estimate of SN rates for all types as a function of redshift has been proposed and
accepted a three years SN search with the VST telescope. To help planning an optimal strategy for the search, we
have developed a simulation tool used to predict the numbers of Supernovae of different types which are expected
to be discovered in a magnitude-limited survey. In our simulation a most important ingredient has been the
determination of the $K$-$correction$ as function of redshift for every SN types and filters set. We estimate to
discover $\sim$300 SNe in the course of this search. This number, above a given threshold, is constant even if
the frequency of observation increases. Moreover we find that to high redshift the infrared band become more
important.}

\section{Introduction}

Supernovae explosion rates provide important information on the evolutionary scenario of SN progenitors, the
star formation history of the Universe and ISM physic (See
Refs~\refcite{madau},~\refcite{Sadat},~\refcite{Dahl},~\refcite{Kob},~\refcite{Sull},~\refcite{Cal}). In recent
years many SN surveys have been conducted aimed to the detection of SN candidates to measure the SN rates as
function of redshift. Most of them are designed to search SNe for their use as cosmological probes
(Refs~\refcite{pain},~\refcite{pain1},~\refcite{tonry}) and, as a consequence, they are strongly biased towards
type Ia and they provide no information on the evolution of core-collapse SN rates. Recently we completed the
first phase of a long term project aimed to measure the rate at intermediate redshift for all SN types
(Ref~\refcite{cap}) which allow to derive the first estimate of the core-collapse SNe rate at redshift $z\sim
0.3$. The next step we have proposed is to perform a more efficient search at VST and to optimize the search
strategy we have written a simulation tool able to predict the numbers of SNe of different types which are
aspected to be discover in a magnitude-limited survey.

\section{Simulation Tool}
Our simulation is based on several assumptions: 1) galaxy luminosity density and rate evolution with redshift
are described by a power law (see Ref ~\refcite{cap}); 2) the geometry of Universe is flat with
$H_{0}=75kms^{-1}$, $\Omega_{M}=0.3$ (see Ref ~\refcite{hogg}); 3) the light curves for different SNe types in
the SN rest frame do not change with redshift. We have created a catalogue of SN spectra extended from the UV to
the infrared using spectra taken from IUE, HST and Asiago archives. The first step has been to join the
different spectral ranges obtaining homogeneus, extended spectra. With redshift the observed SNe spectrum is
stretched and shifted toward longer wavelengths and the SN apparent magnitude is different from that measured in
the SN rest frame. Using our spectral library we estimated the K-correction as a function of redshift for
different SN types and filters combination. In particular we have estimated K-correction for the Sloan filter
set. Besides we have determined the optimal filter band to maximize the number of detectable SNe. Finally, using
the control time method, we have determined the temporal interval between observations which optimize the search
efficiency.
\section{Conclusions}
We have computed the expected SN counts as a function of redshift by summing the contribution of individual SN
types. The number of SNe expected during the three years of our VST SN search is $\sim$300, a good statistics to
improve the SN rate. Because of the lack of UV SN observations at high redshift our K-correction estimates are
uncertain. On the other hand at this distance optical search should be replaced by infrared ones because the
optical emission of high redshift SNe is very low. Another useful result is, at different redshift, the limit on
the frequency of observations over which there is no gain as far as concern the number of SN detection. We aim
to extend the scope of our simulation tool to the planning of SN search with a generic observing facilities. In
particular we are interested to the simulation of a SN search with the new wide field infrared facility VISTA.

%%%%%%%%%%%%%%%%%%%%%%%%%%%%%%%%%%%%%%%%%%%%%%%%%%%%%%%%%%%%%%%%%%%%%%%
%Other ways of embedding figures!
%
%\begin{figure}[t]
%\centerline{\psfig{file=procs-fig1.eps,width=5cm}}
%\vspace*{8pt}
%\caption{A schematic illustration of dissociative recombination. The
%direct mechanism, 4m$^2_\pi$ is initiated when the
%molecular ion S$_{\rm L}$ captures an electron with kinetic energy.}
%\end{figure}
%%%%%%%%%%%%%%%%%%%%%%%%%%%%%%%%%%%%%%%%%%%%%%%%%%%%%%%%%%%%%%%%%%%%%%%

%%%%%%%%%%%%%%%%%%%%%%%%%%%%%%%%%%%%%%%%%%%%%%%%%%%%%%%%%%%%%%%%%%%%%%%
%
%Use this if your figures are put in a subdirectory having the same
%name as the main latex file, ie:
%
%      ws-procs975x65/procs-fig1.eps
%      ws-procs975x65/procs-fig2.eps
%      ws-procs975x65/procs-fig3.eps
%      ws-procs975x65/procs-fig4.eps
%      etc.
%
%\begin{figure}[htbp] %ORIGINAL SIZE: width=1.4TRUEIN; height=1.5TRUEIN
%\figurebox{}{}{procf1} %100 percent
%\caption{Labeled tree {\it T}.}
%\end{figure}
%
%%%%%%%%%%%%%%%%%%%%%%%%%%%%%%%%%%%%%%%%%%%%%%%%%%%%%%%%%%%%%%%%%%%%%%

\end{document}